\documentclass[aps,pre,showpacs,superscriptaddress]{revtex4}

\usepackage{graphicx}

\usepackage{amsmath}

\begin{document}

\title{Soliton solutions of 3D Gross-Pitaevskii equation by a potential control method}

\author{Renato Fedele}
\email{renato.fedele@na.infn.it} \affiliation{Dipartimento di
Scienze Fisiche, Universit\`a Federico II and INFN, Complesso
Universitario di M. S. Angelo, Via Cintia, 80126 Napoli, Italy}

\author{Bengt Eliasson}
\email{bengt@tp4.rub.de} \affiliation{Department of Physics,
Ume{\aa} University, SE-90 187 Ume{\aa}, Sweden}
\affiliation{Institut f\"ur Theoretische Physik IV and Centre for
Nonlinear Physics, Fakult\"at f\"ur Physik und Astronomie,
Ruhr--Universit\"at Bochum, D-44780 Bochum, Germany}

\author{Fernando Haas}
\email{ferhaas@hotmail.com} \affiliation{Institut f\"ur
Theoretische Physik IV and Centre for Nonlinear Physics,
Fakult\"at f\"ur Physik und Astronomie, Ruhr--Universit\"at
Bochum, D-44780 Bochum, Germany}

\author{Padma Kant Shukla}
\email{ps@tp4.rub.de} \affiliation{Institut f\"ur Theoretische
Physik IV and Centre for Nonlinear Physics, Fakult\"at f\"ur
Physik und Astronomie, Ruhr--Universit\"at Bochum, D-44780 Bochum,
Germany} \affiliation{Scottish Universities Physics Alliance
(SUPA), Department of Physics, University of Strathclyde, Glasgow
G4 ONG, United Kingdom}

\author{Dusan Jovanovi\'c}
\email{djovanov@phy.bg.ac.yu} \affiliation{Institute of Physics,
P. O. Box 57, 11001 Belgrade, Serbia}

\author{Sergio De Nicola}
\email{s.denicola@cib.na.cnr.it} \affiliation{Instituto di
Cibernetica ``E Caianiello''  del CNR Comprensorio  ``A.
Olivetti'', Via Campi Flegrei 34, I-80078 Pozzuoli (Na), Italy }

\received{8 July 2009}

\begin{abstract}
We present a class of three-dimensional solitary waves solutions
of the Gross-Pitaevskii (GP) equation, which governs the dynamics
of Bose-Einstein condensates (BECs). By imposing an external
controlling potential, a desired time-dependent shape of the
localized BEC excitation is obtained. The stability of some
obtained localized solutions is checked by solving the
time-dependent GP equation numerically with analytic solutions as
initial conditions. The analytic solutions can be used to design
external potentials to control the localized BECs in experiment.
\end{abstract}


\maketitle

\section{Introduction}

Solitons are nonlinear localized wave packets sustained by the
balance between wave dispersion and medium nonlinearity. Solitons
propagate over large distances without changing their shape
\cite{Sulem99,Whitham99,Dau06}. Solitary waves or solitons have
been observed in several areas of physics including fluids,
plasmas, optics, biology, and condensed matters (e.g.
Bose-Einstein condensates). Many types of solitons have been
studied, starting with classical examples found in integrable
models, such as the Korteweg-de Vries, sine-Gordon, Toda-lattice,
and nonlinear Schr\"odinger equations, and their non-integrable
extensions. Solitons are robust against collisions due to the
integrability of the underlying equations.

Recent observations of matter-wave solitons
\cite{r1a,r1b,r1c,r1d,r2a,r2b,r2c,r2d} have been among the most
groundbreaking achievements in the burgeoning fields of
Bose-Einstein condensation (BEC) of dilute atomic gases. In the
latter, bosonic atoms below a certain temperature suddenly develop
in the lowest quantum mechanical state. The balance between the
spatial dispersion of matter waves and repulsive or attractive
atomic interactions in Bose-Einstein condensates (BECs) ensures
the existence of dark \cite{r1a,r1b,r1c,r1d} or bright
\cite{r2a,r2b,r2c,r2d} solitons, respectively.  A dark (bright)
matter wave soliton is a localized BEC having a minimum (maximum)
condensate density at the center. However, if the atomic
condensate is embedded into  a periodic potential created by
standing light waves, i.e.  optical lattice \cite{optical}, there
exists possibility of reversing the matter wave group dispersion
sign (e.g. from positive group dispersion to negative group
dispersion), and of the possible observation of the bright
matter-wave soliton in the BECs with repulsive inter-atomic
interaction. The concept of the dispersion control by periodic
potentials is also well known in solid state physics \cite{solid}
and a very active topic of research in nonlinear optics
\cite{optics}. The dynamics and stability of the matter wave
solitons in the BECs is governed by the nonlinear Schr\"odinger
equation \cite{Sulem99}, known as the Gross-Pitaevskii (GP)
equation \cite{GP,GP1} in the context of the BECs. For BECs with
positive (repulsive) interparticle interactions, dark solitons
with locally depleted density have been studied theoretically
\cite{Jackson98} and have been observed in many experiments
\cite{r1a,r1b,r1c,r1d,r2a,r2b,r2c,r2d}. Numerical and theoretical
studies revealed the dynamics and stability the BECs in a
magneto-trap both for repulsive and attractive interactions in a
limited parameter range in spherically symmetric trap
\cite{Ruprecht95}. The stabilization and controlling of the BECs
in asymmetric traps have been investigated by considering the
time-dependent solutions of the GP equation \cite{Garcia98}.
Stable condensates with a limit number of atoms of $^7$Li with an
attractive interaction have been observed in a magnetically
trapped gas \cite{Bradley97}. The formation of matter-wave
solitons \cite{r1a,r1b} and trains of solitons \cite{r2a,r2b} have
been observed in BECs of $^7$Li atoms that are confined in a
quasi-one-dimensional trap and magnetically tuned from repulsive
to attractive interactions. The solitons are predicted to either
collapse or explode, depending on the parameters of the BECs and
on the confining or repulsive potential \cite{Carr02}. The
collective collapse \cite{Sackett99,Donley01} and explosion
\cite{Donley01} of BECs with attractive interactions have been
observed in experiments. Several theoretical and experimental
studies of coherent matter waves are contained in Ref.
\cite{JPhysB}. Wang {\it et al.} \cite{Wang03} have presented the
analytical dark and bright solitons of the one-dimensional GP
equation with a confining potential. The generation of matter wave
dark and bright solitons in a prescribed external potential for
confining the BECs have been investigated with a periodically
varying nonlinear coefficient \cite{Kevrekidis03,Saito03}.  A
tight transverse trap with a gradually varying local frequancy
along the longitudinal direction induces an effective potential
for one-dimensional solutions in a self-attracted BEC
\cite{DeNicola}. The propagation of a dark soliton in a quasi-1D
BEC in the presence of a random potential has been studied by
Bilas and Pavloff \cite{Bilas05}. Furthermore, there is a recent
theoretical study of exact one-dimensional solitary wave solutions
in a radially confining potential \cite{Atre06}.  The dynamics of
the one-dimensional bright matter wave soliton in a lattice
potential has been studied by Poletti {\it et al.}
\cite{Poletti08}. Two-dimensional dark solitons to the nonlinear
Schr\"odinger equation are numerically created by two processes to
show their robustness (e.g. stable against the head-on collision).
Stellmer {\it et al.} \cite{Stellmer08} present experimental data
exhibiting the head-on collision of dark solitons generated in an
elongated BEC. Experiments do not report discernible interaction
among solitons, demonstrating the fundamental theoretical concepts
of solitons as quasiparticles.

In this paper, we theoretically study possibility of controlling
the time-dependent dynamics of the BECs in three-dimensions with a
carefully spatially shaped, time-dependent controlling potential.
This idea is theoretically supported by our recent mathematical
investigations \cite{GPE1} and on the improvement of the
recently-proposed 'controlling potential method' \cite{Fedele06}.
From the experimental point of view, this idea could be realized
by techniques involving lithographically designed circuit patterns
that provide electromagnetic guides and microtraps for ultracold
systems of atoms in BEC experiments \cite{Forthagh98}, and by
optically induced ``exotic'' potentials \cite{Grimm00}. The
stability of the obtained analytic solutions are investigated
numerically by direct integration of the time-dependent GP
equation.

\section{Theory}

The dynamics of BECs is in a non-uniform potential is governed by
the three-dimensional Gross-Pitaevskii equation \cite{GP,GP1}
\begin{equation}
i\hbar \frac{\partial \Psi}{\partial t}=-\frac{\hbar^2}{2
m}\nabla^2\Psi+g N |\Psi|^2\Psi+V_{\rm ext}({\bf r},t)\Psi,
\label{eq1}
\end{equation}
where $m$ is the atomic mass, and $V_{\rm ext}$ is the external
potential, $g=4\pi \hbar^2 a/m$ where $a$ is the short-range
scattering length, which can be either positive or negative,
giving rise to either repulsive or attractive interactions, and
$N$ is the number of atoms in the BECs. Typical parameters values
where solitons have been observed in BECs of $^7$Li atoms are
$N=10^4$--$10^5$  at a temperature of $1$--$10\,{\rm \mu K}$ and a
magnetic field $\sim 400$--$600\,\mathrm{G}$, leading to a small
scattering length of $a \approx -0.2\,\mathrm{nm}$ \cite{r1a,r2b}.
In Eq. (\ref{eq1}), $\Psi$ is normalized such that $\int
|\Psi|^2\,d^3r=1$. Equation (\ref{eq1}) can be cast into the
dimensionless form
\begin{equation}
i\frac{\partial \psi}{\partial
t}=-\frac{1}{2}\nabla^2\psi+\widetilde{g} |\psi|^2\psi+U_{\rm
ext}({\bf r},t)\psi, \label{eq2}
\end{equation}
where the time $t$ is normalized by $t_0=(4\pi)^2 m a^2
N^2/\hbar$, space ${\bf r}$ by $r_0=4\pi |a| N$, the external
potential $U_{\bf ext}$ by $\hbar^2/(4\pi)^2 m a^2 N^2$,  and
$\psi=r_0^{3/2}\Psi$, so that $\int |\psi|^2\,d^3r=1$ in the
normalized spatial variables. With this normalization, we have
$\widetilde{g}=+1$ for $a>0$ and $\widetilde{g}=-1$ for $a<0$. For
typical experimental values $N=6000$, $|a|=0.2\,\mathrm{nm}$ and
$m=1.17\times10^{-26}\,\mathrm{kg}$ ($^7$Li) \cite{r2b}, we would
have $r_0=16\,\mathrm{\mu m}$ and $t_0=0.028\,\mathrm{s}$. Our
goal is to design the external potential to obtain a desired
time-dependent shape of the solution.

We concentrate on a sub-class of solutions, in which the GP
equation can be separated into one linear, two-dimensional
equation, and one nonlinear, one-dimensional equation. In doing
so, we first make a decomposition of the external potential
according to \cite{GPE1}
\begin{equation}
  U_{\rm ext}({\bf r},t)=U_\perp({\bf r}_\perp,t)+U_z({\bf r}_\perp,z,t),
\label{eq3}
\end{equation}
where ${\bf r}_\perp$ is the position vector perpendicular to the
$z$ direction, and make the ansatz that the solution can be
separated as
\begin{equation}
  \psi({\bf r},t)=\psi_\perp({\bf r}_\perp,t)\psi_z(z,t),
\label{eq4}
\end{equation}
where the normalization conditions
$\int|\psi_\perp|^2\,d^2r_\perp=1$ and $\int |\psi_z|^2\,dz=1$ are
imposed.

Inserting Eqs. (\ref{eq3}) and (\ref{eq4}) into Eq.  (\ref{eq2}), and reordering the terms, we have
\begin{equation}
\begin{split}
 &\psi_z\left[
i\frac{\partial \psi_\perp}{\partial
t}+\frac{1}{2}\nabla_\perp^2\psi_\perp-U_\perp({\bf
r}_\perp,t)\psi_\perp
 \right]
 \\
 &=-\psi_\perp\left[
i\frac{\partial \psi_z}{\partial
t}+\frac{1}{2}\frac{\partial^2\psi_z}{\partial z^2}- \widetilde{g}
|\psi_\perp|^2|\psi_z|^2\psi_z-U_z({\bf r}_\perp,z,t)\psi_z
 \right],
 \end{split}
\label{eq5}
\end{equation}
where we have denoted $\nabla^2_\perp=\partial^2/\partial
x^2+\partial^2/\partial y^2$. By requiring that $\psi_\perp$
satisfies the linear Schr\"odinger equation
\begin{equation}
i\frac{\partial \psi_\perp}{\partial
t}+\frac{1}{2}\nabla_\perp^2\psi_\perp-U_\perp({\bf
r}_\perp,t)\psi_\perp=0, \label{eq6}
\end{equation}
we obtain the one-dimensional GP equation
\begin{equation}
i\frac{\partial \psi_z}{\partial
t}+\frac{1}{2}\frac{\partial^2\psi_z}{\partial z^2}- \widetilde{g}
|\psi_\perp|^2|\psi_z|^2\psi_z-U_z({\bf r}_\perp,z,t)\psi_z=0.
\label{eq7}
\end{equation}
For the linear Eq. (\ref{eq6}) for $\psi_\perp({\bf r}_\perp,t)$,
we assume a parabolic potential well in the form
\begin{equation}
  U_\perp=\frac{1}{2}K_x(t)x^2+\frac{1}{2}K_y(t)y^2,
 \label{eq41}
\end{equation}
which yields solutions of Eq. (\ref{eq6}) in terms of functions of
the form $\psi_{\perp mn}(x,y,t)=\psi_{xm}(x,t)\psi_{yn}(y,t)$
where $\psi_{\alpha\beta}$ (where $\alpha=x,\,y$ and
$\beta=0,\,1,\,\ldots$) are Hermite-Gauss functions in the form
\begin{equation}
\begin{split}
  &\psi_{\alpha\beta}(\alpha,t)=\frac{\exp[-\alpha^2/4\sigma_\alpha^2(t)]}
  {[2\pi\sigma_\alpha^2(t)2^{2\beta}(\beta!)^2]^{1/4}}
  {\rm H}_\beta\left[\frac{\alpha}{\sqrt{2}\sigma_\alpha(t)}\right]
  \exp\left[i\frac{\gamma_\alpha(t)\alpha^2}{2}+i\phi_{\alpha\beta}(t)\right],
\end{split}
 \label{eq42}
\end{equation}
(normalized so that
$\int_{-\infty}^{\infty}|\psi_{\alpha\beta}(\alpha,t)|^2\,d\alpha=1$)
where ${\rm H}_\beta(\xi)$ are Hermite polynomials of order
$\beta$. The first few Hermite polynomials are listed in Table
\ref{TableI} in the Appendix A. Here the phases are given in terms
of $\sigma_\alpha$ as
\begin{equation}
  \gamma_\alpha(t)=\frac{1}{\sigma_\alpha}\frac{d\sigma_\alpha}{dt},
 \label{eq43}
\end{equation}
and $\phi_{\alpha\beta}(t)=(2\beta+1)\phi_{\alpha 0}(t)$, where
\begin{equation}
  \frac{d\phi_{\alpha 0}(t)}{d t}=-\frac{1}{4\sigma_\alpha^2(t)}.
 \label{eq45}
\end{equation}
We have that $\sigma_\alpha$ is related to $K_\alpha$ by the
Pinney equation \cite{Pinney}
\begin{equation}
  \frac{d^2\sigma_\alpha}{dt^2}+K_\alpha(t)\sigma_\alpha-\frac{1}{4\sigma_\alpha^3}=0.
 \label{eq44}
\end{equation}
Solving for $K_\alpha(t)$ in (\ref{eq44}) and inserting the result
into (\ref{eq41}) we have the potential well
\begin{equation}
  U_\perp=\frac{1}{2}\left(-\frac{1}{\sigma_x}\frac{d^2 \sigma_x}{dt^2}+\frac{1}{4\sigma_x^4}\right)x^2
  +\frac{1}{2}\left(-\frac{1}{\sigma_y}\frac{d^2 \sigma_y}{dt^2}+\frac{1}{4\sigma_y^4}\right)y^2.
 \label{U_perp}
\end{equation}
in terms of $\sigma_x(t)$ and $\sigma_y(t)$. Hence, we have the
possibility to arbitrarily choosing the time-dependent widths
$\sigma_x(t)$ and $\sigma_y(t)$ (and the indices $m$ and $n$) of
our solution, and obtain the potential $U_{perp}$ necessary to
sustain that solution. Now, the solution of Eq. (\ref{eq7}) for
$\psi_z$ must be such that the total $\psi({\bf r},t)$ in
(\ref{eq4}) solves the original GP equation (\ref{eq2}). A special
solution is
\begin{equation}
  \psi_z(z,t)=\left[-\frac{\widetilde{g}\delta_m\delta_n}{4\sigma_x(t)\sigma_y(t)}\right]^{1/2}
  {\rm sech}\left[-\frac{\widetilde{g}\delta_m\delta_n}{2\sigma_x(t)\sigma_y(t)} z\right]
  \exp\left[\frac{i}{2}g(t)z^2+i\theta_0(t)\right],
 \label{z1}
\end{equation}
where phase functions are given by
\begin{equation}
  g=\frac{1}{\sigma_x(t)\sigma_y(t)}\frac{d [\sigma_x(t)\sigma_y(t)]}{dt}.
  \label{z2}
\end{equation}
and
\begin{equation}
 \frac{d\theta_0}{dt}=\frac{1}{8}\left[\frac{\widetilde{g}\delta_m\delta_n}{\sigma_x(t)\sigma_y(t)}\right]^2.
 \label{z3}
\end{equation}
The external potential $U_z$ to sustain this solution is given by
\begin{equation}
  U_z({\bf r}_\perp,z,t)=-\widetilde{g}\left[ |\psi_\perp|^2-\frac{\delta_m\delta_n}{\sigma_x(t)\sigma_y(t)}
  \right]|\psi_z|^2+\frac{1}{2}K(t)z^2,
  \label{z4}
\end{equation}
where
\begin{equation}
  K(t)=-\frac{1}{\sigma_x(t)\sigma_y(t)}\frac{d^2}{dt^2}\left[\sigma_x(t)\sigma_y(t)\right].
 \label{z5}
\end{equation}
The details of the derivation of $\psi_z$ and $U_z$ are given in
the Appendix A. The perpendicular solutions $\psi_\perp$ in
(\ref{eq42}) and parallel solution $\psi_z$ in (\ref{z1}) can now
be used to construct the total, three-dimensional solution $\psi$
in (\ref{eq4}) of the GP equation (\ref{eq2}), and where the
external controlling potential $U_{ext}$ in (\ref{eq3}) is the sum
of $U_\perp$ in (\ref{U_perp}) and $U_z$ in (\ref{z4}).

\section{Analytical and numerical examples}
\begin{figure}[htb]
\centering
\includegraphics[width=8.5cm]{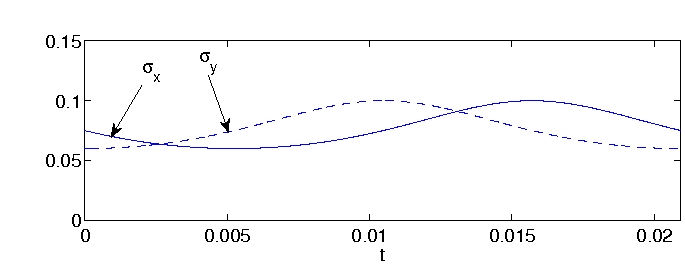}
\caption{The soliton widths $\sigma_x$ (solid line) and $\sigma_y$
(dashed line) as functions of time, during one oscillation period,
for the equilibrium widths $\sigma_{0x}=\sigma_{0y}=0.075$, the
oscillation amplitudes $a_{0x}=a_{0y}=0.25$, and the frequency
$\omega=300$.} \label{Fig:sigma_xy}
\end{figure}
\begin{figure}[htb]
\centering
\includegraphics[width=8.5cm]{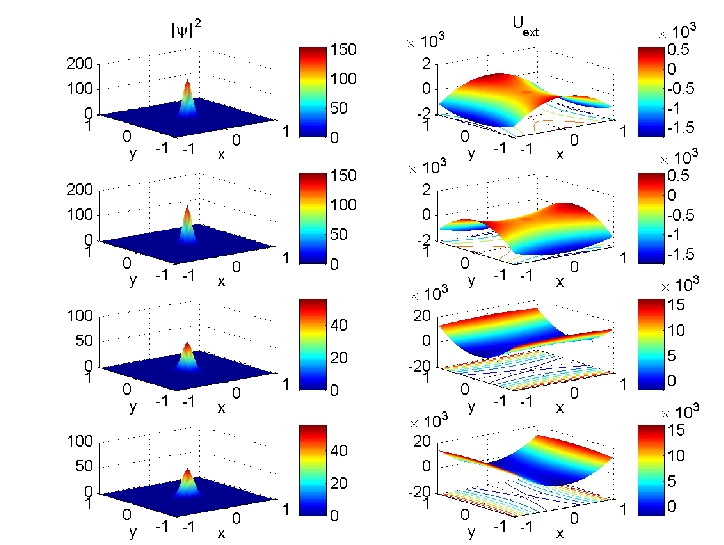}
\caption{The condensate density $|\psi|^2$ (left) and the external
potential $U_{ext}$ (right) in the $xy$ plane at $z=0$, at times
$t=0$, $t=0.25\times 2\pi/\omega$, $t=0.5\times 2\pi/\omega$, and
$t=0.75\times 2\pi/\omega$ (top to bottom rows), for the ground
state $(m,n)=(0,0)$.} \label{Fig:plots_00_0.25}
\end{figure}

\begin{figure}[htb]
\centering
\includegraphics[width=8.5cm]{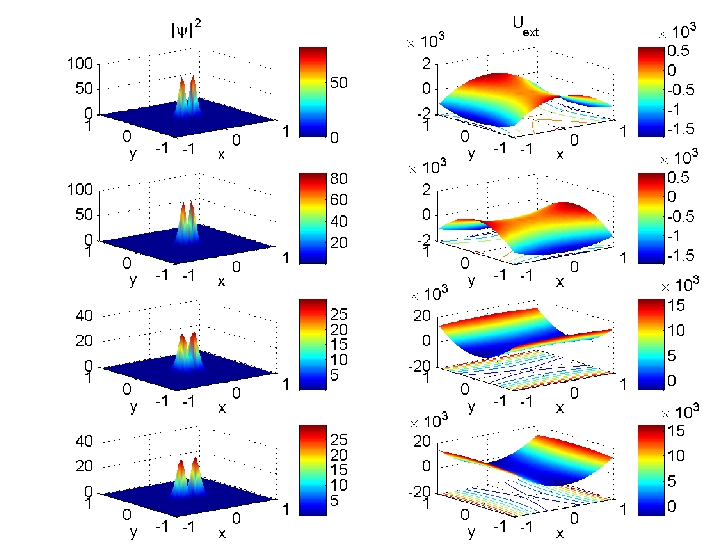}
\caption{The condensate density $|\psi|^2$ (left) and the external
potential $U_{ext}$ (right) in the $xy$ plane at $z=0$, at times
$t=0$, $t=0.25\times 2\pi/\omega$, $t=0.5\times 2\pi/\omega$, and
$t=0.75\times 2\pi/\omega$ (top to bottom rows), for the excited
state $(m,n)=(1,0)$.} \label{Fig:plots_10_0.25}
\end{figure}

\begin{figure}[htb]
\centering
\includegraphics[width=8.5cm]{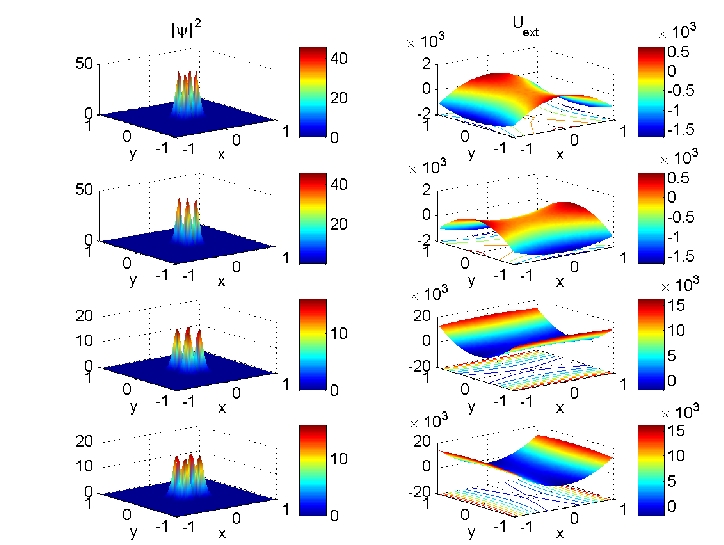}
\caption{The condensate density $|\psi|^2$ (left) and the external
potential $U_{ext}$ (right) in the $xy$ plane at $z=0$, at times
$t=0$, $t=0.25\times 2\pi/\omega$, $t=0.5\times 2\pi/\omega$, and
$t=0.75\times 2\pi/\omega$ (top to bottom rows), for the excited
state $(m,n)=(1,1)$.} \label{Fig:plots_11_0.25}
\end{figure}























We will here exemplify our analytic results with exact
time-dependent solutions of the GP equation. As mentioned above,
we are free to choose arbitrary time-dependencies of the widths
$\sigma_x(t)$ and $\sigma_y(t)$. In Eq. (\ref{eq42}) we also need
the Hermite polynomials ${\rm H}_m$ and in Eqs. (\ref{z1}),
(\ref{z3}) and (\ref{z4}) we need the constants $\delta_m$, which
are given in Table \ref{TableI} in Appendix A, for different
excited states $m$. As an example, we choose $\sigma_x(t)$ and
$\sigma_y(t)$ to be periodic in time, of the form
\begin{equation}
 \sigma_x(t)=\frac{\sigma_{0x}}{1+ a_{0x}\sin(\omega t)},
 \label{sigma_x}
\end{equation}
and
\begin{equation}
  \sigma_y(t)=\frac{\sigma_{0y}}{1+a_{0y}\cos(\omega t)},
  \label{sigma_y}
\end{equation}
where we investigate cases of relatively large amplitude
oscillations with $a_{0x}=a_{0y}=0.25$, and we set
$\sigma_{0x}=\sigma_{0y}=0.075$ and $\omega=300$. In dimensional
units with $r_0=16\,\mathrm{\mu m}$ and $t_0=0.028\,\mathrm{s}$,
this corresponds to a typical soliton width of $\sim 0.075\times
16\,\mathrm{\mu m}\approx 1.5\,\mathrm{\mu m}$ and the frequency
$300/(2\pi\times 0.028)\,\mathrm{Hz}=1.7\,\mathrm{kHz}$. The
choice of the functions (\ref{sigma_x}) and (\ref{sigma_y})
describes periodically in time pulsating solutions widths,
illustrated in Fig.~\ref{Fig:sigma_xy}. Small values of $\sigma_x$
and $\sigma_y$ correspond to spatially localized solitary waves,
while large values correspond to wider solitary waves. The
amplitude of the solitary wave also varies so as to keep the total
number of condensates constant, $\int|\psi|d^3 r=1$. In Figs.
\ref{Fig:plots_00_0.25}--\ref{Fig:plots_11_0.25}, we have plotted
the solutions together with the corresponding controlling
potentials in the $xy$-plane at $z=0$, for the ground state
$(m,n)=(0,0)$ and the excited states $(m,n)=(1,0)$ and
$(m,n)=(1,1)$, where $(m,n)$ refers to the orders of the Hermite
polynomials in the perpendicular solution $\psi_{\perp
mn}(x,y,t)$.  The widths are varying such that at time $t=0$, the
solitons are localized and large amplitude while at later times
their amplitudes decrease and their widths increase, first in the
$y$ direction and then in the $x$ direction. We note that the
external potential is sometimes large amplitude and confining, and
sometimes small amplitude and non-confining.

\begin{figure}[htb]
\centering
\includegraphics[width=8.5cm]{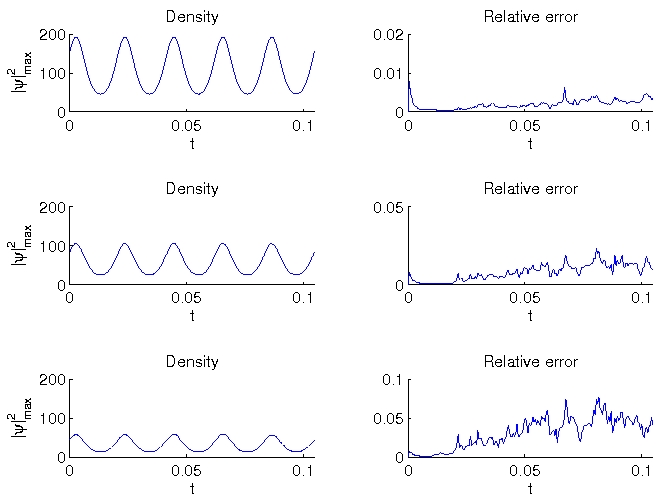}
\caption{Numerical simulation results of the GP equation for the
amplitude $a_{0x}=a_{0y}=0.25$ and $(m,n)=(0,0)$, $(1,0)$ and
$(1,1)$ (top to bottom panels), showing the maximum density (left
column) and the relative deviation of the numerical solution from
the exact analytic solution (right column). }
\label{Fig:stability_small}
\end{figure}

For the analytic results to be observable in experiments, it is
necessary that they are stable. To assess the stability of the
solutions, we have therefore solved the time-dependent GP equation
(2) in three dimensions with the analytic solution as initial
condition at time $t=0$. We used a box length $L_x=L_y=2$ in the
$x$ and $y$ dimensions and $L_z=3$ in the $z$ direction, with
periodic boundary conditions. The spatial derivatives were
approximated with a pseudospectral method, and the time stepping
was performed with the standard 4th-order Runge-Kutta method. The
space was resolved with 64 grid points in the $x$ and $y$
directions and with 200 grid points in the $z$ direction, and the
timestep was taken to be $\Delta t=\pi/300\,000$.

In order to seed any instability in the system, we added random
perturbations in phase to the initial condition of the order 0.01
rad. We then simulated the system for 5 oscillation periods and
measured the maximum density $|\psi|^2_{num, max}$ obtained in the
numerical solution as well as the maximum relative error in
density fluctuations,
$\varepsilon=|(|\psi|^2-|\psi|^2_{num})|_{max}/|\psi|^2_{max}$, as
a function of time, and plotted the results in Fig.
\ref{Fig:stability_small}, for the ground state $(m,n)=(0,0)$ and
the excited states $(m,n)=(1,0)$ and $(m,n)=(1,1)$. We see that
the numerical solution follows almost exactly the analytic
solution, without increasing substantially throughout the
simulation. The ground state $(m,n)=(0,0)$ shows a relative error
less than 1 \% throughout the simulation while the excited state
$(m,n)=(1,1)$ shows a somewhat larger error but still less than 10
\%. Hence the controlled BEC seems to be stable enough to be
observed in experiment.

\section{Summary and conclusions}
In summary, we have presented a class of three-dimensional
solitary waves solutions of the Gross-Pitaevskii (GP) equation,
which governs the dynamics of Bose-Einstein condensates. This has
been done on the basis of our recent mathematical investigations
\cite{GPE1} and improving the formulation of the recently-proposed
controlling potential method \cite{Fedele06}. By imposing an
external controlling potential, a desired time-dependent shape of
the localized BECs is obtained. The stability of the exact
solutions were checked with direct simulations of the
time-dependent, three-dimensional GP equation. Our  simulations
show that the localized condensates are stable with respect to
perturbed initial conditions.  We propose that our findings could
be tested experimentally by techniques involving lithographically
designed circuit patterns that provide electromagnetic guides and
microtraps for ultracold systems of atoms in BEC experiments
\cite{Forthagh98}, and by optically induced ``exotic'' potentials
\cite{Grimm00}. Furthermore, numerical simulations \cite{Maris05}
reveal that soliton emission from a BEC can be controlled by a
shallow optical dipole trap. Here, the emission of matter wave
bursts is triggered by spatial variation of the scattering length
along the trapping axis. The motion of the 1D dark soliton can
also be controlled by means of periodic potentials in optical
lattices \cite{Theo05}. Finally, it should be noted that repulsive
BECs confined by an optical lattice and a parabolic magnetic trap
can appear in the form of vortices \cite{Kev03} as well. Computer
simulations \cite{Weiler08} reveal the condensation of a finite
temperature Bose gas in the form of a single vortex.  Interactions
of solitary waves and vortex rings in a cylindrically controlled
BECs exhibit robustness during head-on collisions
\cite{Komineas05}.

\appendix

\section{Details in the derivation of the parallel solution $\psi_z$ and the potential $V_z$}

We here present a special solution of Eq. (\ref{eq7}) for
$\psi_z$, such that the total $\psi({\bf r},t)=\psi_\perp({\bf
r_\perp},t)\psi_z(z,t)$ solves the original GP equation
(\ref{eq2}). Multiplying Eq. (\ref{eq7}) by $|\psi_\perp|^2$ and
integrating over ${\bf r}_\perp$ space, we obtain
\begin{equation}
i\frac{\partial \psi_z}{\partial
t}+\frac{1}{2}\frac{\partial^2\psi_z}{\partial z^2}-
q_0(t)|\psi_z|^2\psi_z-V(z,t)\psi_z=0, \label{eq8}
\end{equation}
where $V(z,t)=\int U_z({\bf
r}_\perp,z,t)|\psi_\perp|^2\,d^2r_\perp$ and
\begin{equation}
  q_0(t)=\widetilde{g}\int|\psi_\perp|^4\,dx\,dy=\frac{\widetilde{g}\delta_m\delta_n}{\sigma_x(t)\sigma_y(t)},
 \label{eq46}
\end{equation}
where the numerical factor
\begin{equation}
\delta_m =\frac{1}{\sqrt{2}\pi
2^{2m}(m!)^2}\int_{-\infty}^{\infty}\exp(-2\xi^2)|{\rm
H}_m(\xi)|^4\,d\xi
 \label{eq47}
\end{equation}
is evaluated for given values of $m$. The first few Hermite
polynomials and values of $\delta_m$ are listed in Table
\ref{TableI}.
\begin{table}[htb]
\begin{tabular}{|c|c|c|}
\hline
$m$ & $H_m(\xi)$ & $\delta_m $\\
\hline
$0$ & $1$ & $1/(2\sqrt{\pi})$\\
$1$ & $2\xi$ & $3/(8\sqrt{\pi})$\\
$2$ & $4\xi^2-2$ & $41/(128\sqrt{\pi})$\\
\hline
\end{tabular}
\caption{The Hermite polynomial $H_m(\xi)$ and the value of
$\delta_m$ for $m=0$, $1$, and $2$.} \label{TableI}
\end{table}

Equations (\ref{eq7}) and (\ref{eq8}) should give the same
solution for $\psi_z$, and this imposes restrictions on the
external potential. Subtracting Eq. (\ref{eq7}) from Eq.
(\ref{eq8}) we obtain the compatibility relation
\begin{equation}
U_z({\bf
r}_\perp,z,t)=-\widetilde{g}|\psi_\perp|^2|\psi_z|^2+q_0(t)|\psi_z|^2+V.
  \label{eq10}
\end{equation}
We will consider the special case where the external potential for
the one-dimensional GP equation (\ref{eq8}) has the form $V(z,t)=
({1}/{2})K(t)z^2$ so that
\begin{equation}
i\frac{\partial \psi_z}{\partial
t}+\frac{1}{2}\frac{\partial^2\psi_z}{\partial z^2}-
q_0(t)|\psi_z|^2\psi_z-\frac{1}{2}K(t)z^2\psi_z=0. \label{eq11}
\end{equation}
Using the Madelung fluid ansatz \cite{Fedele02}
$\psi_z=\sqrt{\rho(z,t)}\exp[i\theta(z,t)]$, we obtain after some
manipulations [cf. Eq. (11) of Ref. \cite{FedeleSchamel02}]
\begin{align}
\begin{split}
 &-\rho\frac{\partial v}{\partial t}+v\frac{\partial \rho}{\partial t}
 + 2\left[c_0(t)-\int^z\frac{\partial v}{\partial t}dz
 \right]\frac{\partial \rho}{\partial z}
 -\left(\rho\frac{\partial U}{\partial z}+2U\frac{\partial \rho}{\partial z}
 \right)+\frac{1}{4}\frac{\partial^3\rho}{\partial z^3}=0,
 \end{split}
 \label{eq12}
 \intertext{and}
 & \frac{\partial \rho}{\partial t}+\frac{\partial}{\partial z}(\rho
 v)=0,
 \label{eq13}
\intertext{where}
 & v=\frac{\partial \theta}{\partial z}.
 \label{eq14}
\end{align}
In Eq. (\ref{eq12}), $c_0(t)$ is an arbitrary function of $t$ to
be determined below and we have denoted
$U=q_0(t)|\psi_z|^2+V(z,t)=q_0(t)\rho+V(z,t)$. In order to find
solitary wave solutions of Eqs. (\ref{eq12}) and (\ref{eq13}), we
now assume that
\begin{equation}
   v(z,t)=g(t)z,
 \label{eq15}
\end{equation}
where $g(t)$ (not to be confused with the coupling parameter of
the GP equation) is a function to be determined, and we have from
Eq. (\ref{eq14}) that
\begin{equation}
  \theta(z,t)=\frac{1}{2}g(t)z^2+\theta_0(t),
 \label{eq16}
\end{equation}
where $\theta_0(t)$ is an arbitrary function of $t$. Even if a
time-dependent phase is included in the transverse solutions
$\psi_\perp$ below, it is necessary to keep $\theta_0$, since the
parallel solutions are required to satisfy a separate equation.
Using Eq. (A9), together with the relations $\partial v/\partial
t=g'(t)z$ and $\int (\partial v/\partial t)dz=(1/2)g'(t)z^2$
(where the primes denote derivatives) in Eqs.  (\ref{eq12}) and
(\ref{eq13}), we have
\begin{align}
\begin{split}
&-g'(t)z\rho+g(t)z\frac{\partial \rho}{\partial t}+ 2\left[
c_0(t)-\frac{1}{2}g'(t)z^2 \right]\frac{\partial \rho}{\partial z}
\\
&-3q_0 \rho\frac{\partial \rho}{\partial z}-\left(
Kz\rho+Kz^2\frac{\partial \rho}{\partial z}\right) +
\frac{1}{4}\frac{\partial^3\rho}{\partial z^3}=0
\end{split}
 \label{eq17}
\intertext{and} &\frac{\partial \rho}{\partial
t}+g(t)z\frac{\partial \rho}{\partial z}+g(t)\rho=0,
 \label{eq18}
\end{align}
respectively. Multiplying Eq. (\ref{eq18}) by $g(t)z$ and
subtracting the result from Eq. (\ref{eq17}), and reordering the
terms, we have
\begin{equation}
 -z[g'(t)+g^2(t)+K(t)]\left(z\frac{\partial \rho}{\partial z}+\rho\right)
 +2c_0(t)\frac{\partial\rho}{\partial z}
 -3q_0\rho\frac{\partial \rho}{\partial
 z}+\frac{1}{4}\frac{\partial^3\rho}{\partial z^3}=0.
 \label{eq19}
\end{equation}
If $g(t)$ obeys the Riccati equation
\begin{equation}
  g'(t)+g^2(t)+K(t)=0,
 \label{eq20}
\end{equation}
then Eq. (\ref{eq19}) simplifies to
\begin{equation}
 2c_0(t)\frac{\partial\rho}{\partial z}
 -3q_0\rho\frac{\partial \rho}{\partial
 z}+\frac{1}{4}\frac{\partial^3\rho}{\partial z^3}=0.
 \label{eq21}
\end{equation}
We now introduce the change of variables $\xi =G(t)z+R(t)$ and
$\tau=t$, or
\begin{equation}
z =[\xi-R(\tau)]/G(\tau)
 \label{eq24}
\end{equation}
and
\begin{equation}
t=\tau,
 \label{eq25}
\end{equation}
which can be introduced into Eqs. (\ref{eq19}) and (\ref{eq18}) to
obtain
\begin{align}
&2c_0 G\frac{\partial \rho}{\partial \xi}-3q_0 G
\rho\frac{\partial \rho}{\partial
\xi}+\frac{1}{4}G^3\frac{\partial^3\rho}{\partial\xi^3}=0,
 \label{eq26}
\intertext{and}
&(\xi-R)\left[\frac{G'(\tau)}{G}+g(\tau)\right]\frac{\partial\rho}{\partial\xi}
+R'(\tau)\frac{\partial\rho}{\partial\xi}+g\rho+\frac{\partial\rho}{\partial\tau}=0,
 \label{eq27}
\end{align}
respectively. Choosing
\begin{equation}
G'(\tau)/G(\tau)+g(\tau)=0,
 \label{eq28}
\end{equation}
and $R=$constant ($R=0$ without loss of generality), Eq.
(\ref{eq27}) becomes
\begin{align}
\frac{\partial \rho}{\partial \tau}+g\rho=0.
 \label{eq29}
\end{align}
We now look for a solution in the form
$\rho(\xi,\tau)=A(\tau)F(\xi)$. With this ansatz, Eqs.
(\ref{eq26}) and (\ref{eq29}) can be written as
\begin{align}
&
2c_0(\tau)F'(\xi)-3q_0(\tau)A(\tau)F(\tau)F'(\xi)+\frac{1}{4}G^2(\tau)F'''(\xi)=0,
 \label{eq30}
\intertext{and} &A'(\tau)+g(\tau)A(\tau)=0,
 \label{eq31}
\end{align}
respectively. For consistency, Eq. (\ref{eq30}) for $F(\tau)$
should be reduced to an equation with the coefficients independent
of $\tau$, i.e. $c_0(\tau)$ and $q_0(\tau)A(\tau)$ must be
proportional to $G^2(\tau)$. Equations (\ref{eq28}) and
(\ref{eq31}) also imply that $A$ is proportional to $G$.
Integrating Eq. (\ref{eq28}) as
\begin{equation}
 G(\tau)=G_0\exp\left(-\int_0^\tau g(s)\,ds\right),
 \label{eq32}
\end{equation}
where we without loss of generality can choose $G_0=1$, we then
have
\begin{equation}
  A(\tau)=A_0G(\tau),
 \label{eq33}
\end{equation}
\begin{equation}
  c_0(\tau)=C_0G^2(\tau),
 \label{eq34}
\end{equation}
and
\begin{equation}
  q_0(\tau)=Q_0 G(\tau),
 \label{eq35}
\end{equation}
so that Eq. (\ref{eq30}) takes the form (after eliminating the
common exponential factor)
\begin{equation}
 2C_0 F'(\xi)-3A_0Q_0 F'(\xi)F(\xi)+\frac{1}{4}F'''(\xi)=0.
 \label{eq37}
\end{equation}
Equation (\ref{eq37}), which is the time-independent Korteweg-de
Vries equation, admits solitary wave solutions for $C_0<0$ and
$A_0Q_0<0$ in the form
\begin{equation}
  F(\xi)=\frac{2C_0}{A_0 Q_0}{\rm sech}^2\left(\frac{\xi}{\Delta}\right),
 \label{eq38}
\end{equation}
where $\Delta=1/\sqrt{2|C_0|}$. It follows that
\begin{equation}
  \rho(z,t)=\frac{1}{2\Delta}G(t){\rm
  sech}^2\left[\frac{G(t)z}{\Delta}\right].
 \label{eq39}
\end{equation}
with $A_0 \neq 0$, and where we have chosen $Q_0=-2/\Delta$ so
that $q_0(t)=-(2/\Delta)G$ to ensure that
$\int_{-\infty}^{\infty}|\psi_z|^2\,dz=\int_{-\infty}^{\infty}\rho\,dz=1$.
From (\ref{eq46}) we then have
\begin{equation}
  -\frac{2}{\Delta}G=\frac{\widetilde{g}\delta_m\delta_n}{\sigma_x(t)\sigma_y(t)}.
\end{equation}
Since $G=1$ at $t=0$, we find the width
\begin{equation}
  \Delta=-\frac{2\sigma_x(0)\sigma_y(0)}{\widetilde{g} \delta_m\delta_n},
  \label{tilde_g}
\end{equation}
so that
\begin{equation}
  G=\frac{\sigma_x(0)\sigma_y(0)}{\sigma_x(t)\sigma_y(t)},
  \label{G}
\end{equation}
and
\begin{equation}
  g=\frac{1}{\sigma_x(t)\sigma_y(t)}\frac{d [\sigma_x(t)\sigma_y(t)]}{dt}.
  \label{g}
\end{equation}
Finally we have
\begin{equation}
  \psi_z(z,t)=\left[-\frac{\widetilde{g}\delta_m\delta_n}{4\sigma_x(t)\sigma_y(t)}\right]^{1/2}
  {\rm sech}\left[-\frac{\widetilde{g}\delta_m\delta_n}{2\sigma_x(t)\sigma_y(t)} z\right]
  \exp\left[\frac{i}{2}g(t)z^2+i\theta_0(t)\right],
 \label{eq40}
\end{equation}
where $g$ is given by (\ref{g}). To obtain an expression for
$\theta_0(t)$ in terms of $\sigma_x$ and $\sigma_y$, we use Eq.
(\ref{eq40}) to calculate $U(z,t)$, $\rho(z,t)$ and $v(z,t)$,
substitute the result into Eq. (\ref{eq12}), taking into account
that $g = -\dot{G}/G$. This gives $d\theta_{0}/dt = - c_0$, which
together with (\ref{eq34}), (\ref{tilde_g}) and (\ref{G}) yields
the result
\begin{equation}
  \frac{d\theta_0}{dt}=\frac{1}{8}\left[\frac{\widetilde{g}\delta_m\delta_n}{\sigma_x(t)\sigma_y(t)}\right]^2.
\end{equation}
Using (\ref{eq46}) into (\ref{eq10}), we obtain the external
potential
\begin{equation}
  U_z({\bf r}_\perp,z,t)=-\widetilde{g}\left[ |\psi_\perp|^2-\frac{\delta_m\delta_n}{\sigma_x(t)\sigma_y(t)}
  \right]|\psi_z|^2+\frac{1}{2}K(t)z^2,
  \label{U_z}
\end{equation}
where $K$ is found from the Riccati equation (\ref{eq20}) and
Eq.~(\ref{g}) as
\begin{equation}
  K(t)=-\frac{1}{\sigma_x(t)\sigma_y(t)}\frac{d^2}{dt^2}\left[\sigma_x(t)\sigma_y(t)\right].
 \label{K}
\end{equation}

\acknowledgments This work was partially supported by the
Alexander von Humboldt Foundation, by the Deutsche
Forschungsgemeinschaft (Bonn, Germany) through the project
SH21/3-1 of the Research Unit 1048, and by the Swedish Research
Council (VR). B.~E.~acknowledges the support and hospitality of
Universit\`a Federico II, where part of this work was carried out.

\end{document}